\documentclass[prl,11pt,superscriptaddress,floatfix,nofootinbib,notitlepage,longbibliography]{revtex4-1}
\usepackage{graphicx}
\usepackage[colorlinks=true,citecolor=blue]{hyperref}
\usepackage{amsmath}
\usepackage{xcolor}
\usepackage{amssymb}
\usepackage{soul}
\graphicspath{{figures/}}
\begin{document}
\title {Atomically Reconfigurable Single-Molecule Optoelectronics} 

\author{Atif Ghafoor}

\affiliation{Department of Physics, Nanoscience Center, 
University of Jyväskyl\"a, FI-40014 University of Jyväskyl\"a, Finland}

\author{Santeri Neuvonen}

\affiliation{Department of Chemistry, Nanoscience Center, 
University of Jyväskyl\"a, FI-40014 University of Jyväskyl\"a, Finland}

\author{Thinh Tran }
\affiliation{Department of Physics, Nanoscience Center, 
University of Jyväskyl\"a, FI-40014 University of Jyväskyl\"a, Finland}

\author{Oscar Moreno Segura }
\affiliation{Department of Physics, Nanoscience Center, 
University of Jyväskyl\"a, FI-40014 University of Jyväskyl\"a, Finland}

\author{Yitao Sun}
\affiliation{Aalto University, Department of Applied Physics, 00076 Aalto, Finland}

\author{Yaroslav Pavlyukh}
\affiliation{Institute of Theoretical Physics, Faculty of Fundamental Problems of Technology, 
Wroclaw University of Science and Technology, 50-370 Wroclaw, Poland}

\author{Riku Tuovinen}

\affiliation{Department of Physics, Nanoscience Center, 
University of Jyväskyl\"a, FI-40014 University of Jyväskyl\"a, Finland}

\author{Jose L. Lado}

\affiliation{Aalto University, Department of Applied Physics, 00076 Aalto, Finland}

\author{Shawulienu Kezilebieke}
\email{Corresponding authors. Email: kezilebieke.a.shawulienu@jyu.fi}
\affiliation{Department of Physics, Department of Chemistry and Nanoscience Center, 
University of Jyväskyl\"a, FI-40014 University of Jyväskyl\"a, Finland}
 \date{\today}

\begin{abstract}
Deterministic control of excitonic properties is key to advancing nanoscale optoelectronic and quantum technologies and to understanding diverse physical, optical, chemical, and biological phenomena. At the molecular scale, these properties can be tuned through chemical modification, local-environment influence or charge-state manipulation. Yet, direct control of a molecule’s transition dipole moment and its resulting light emission via atomic-scale structural modification has remained elusive. Here, using scanning tunnelling microscopy–induced luminescence, we show that a single structural parameter—the vertical displacement of the central metal atom in a planar phthalocyanine molecule on a decoupling layer—enables active tuning of the transition dipole, allowing either suppression or enhancement of emission. Exploiting this control, we realized a tunable homodimer switchable among three optical states: non-emissive, single-molecule-like emissive, and coupled states exhibiting subradiant and superradiant modes, directly revealing intermolecular dipole–dipole coupling. We further demonstrate a heterodimer in which resonant energy transfer can be turned on or off simply by controlling the acceptor’s transition dipole moment. These findings not only establish atomic-scale displacement as a general strategy for optical molecular switching, but also demonstrate the reconfigurable engineering of excitonic interactions within molecular assemblies.
\end{abstract}
\maketitle
Molecular switches capable of operating at the single-molecule level are central to the development of bottom-up nanotechnologies, offering transformative potential for biosensors, data storage, molecular quantum devices, and optoelectronic applications \cite{feringa2011molecular,vogelsang2009controlling}. One of the most striking ways to modify their optical properties is by switching a molecule between bright and dark states. Typically, chromophores enter dark states either via long-lived, non-emissive triplet states \cite{basche1995direct} or by changes in their redox state that suppress the formation of charge-neutral excited states \cite{zondervan2003photoblinking,kaiser2024gating}. Generally, unwanted and uncontrolled transitions between bright and dark states, commonly observed in molecular fluorescence time traces, are referred to as blinking \cite{rasnik2006nonblinking}. In addition to photobleaching, the formation of irreversible dark states poses a major challenge across many areas of research \cite{vogelsang2008reducing}. To date, no study has demonstrated a strategy to controllably and reversibly switch between bright and dark states at the molecular level through any structural parameter, nor to elucidate the underlying mechanisms of both states within the same molecule.

Scanning probe–based optical spectroscopies provide powerful means to excite and probe bright states of chromophores using either electron or photon excitation \cite{kuhnke2017atomic,yang2020sub,lee2014vibronic,schneider2012plasmonic,qiu2003vibrationally,acsnano.5c04193,zhu2025probing,Doppagne2018-science}. Among these techniques, scanning tunneling microscopy–induced luminescence (STML) is particularly prominent. By exploiting electroluminescence, STML enables the creation of excitons at the single-molecule level and allows direct monitoring of exciton–exciton coupling and energy transfer through the wavelength and intensity of the emitted light \cite{chen2010viewing,Imada2016-va,Zhang2016-yang,Cao2021-ts,hung2024activating,Rai2020-uv,Kong2022-tz}. Pioneering STML experiments have revealed energy transfer within heterogeneous molecular dimers  \cite{Imada2016-va} and demonstrated coherent dipole–dipole coupling in homodimers \cite{Zhang2016-yang}, establishing the foundation of molecular-scale excitonics. Later studies expanded this picture by investigating plasmon–exciton interactions \cite{Imada2017,Zhang2017-yao}, vibrationally assisted photon emission \cite{Doppagne2017-cq}, up-conversion \cite{chen2019spin,luo2024anomalously}, uncovering additional pathways for photon generation \cite{zhang2017electrically} and coherence at the nanoscale \cite{luo2019electrically}. Despite these advances, existing STML studies on chromophores have remained largely observational, revealing excitonic interactions but offering no genuine means of active control. This raises a central question: can molecular switching be achieved through modulation of the transition dipole moment, deterministically controlled by a structural parameter to access bright and dark states? If such a parameter can be realized, could collective optical phenomena in oligomers—such as intermolecular dipole–dipole coupling and energy transfer—be deliberately tuned on demand? Achieving this would transform molecular optoelectronics from a purely observational platform into a reconfigurable one, enabling deterministic control over light–matter interactions at the nanoscale.

In this work, we address this challenge by introducing a controllable structural degree of freedom: the vertical displacement of the central atom in the tin phthalocyanine (SnPc) molecule. We show that upward displacement of the Sn atom enhances the transition dipole moment, enabling bright electroluminescence in the SnPc up configuration. In contrast, lowering the Sn atom suppresses the transition dipole, resulting in negligible emission from the SnPc down configuration. This behavior arises from symmetry breaking and the lifting of orbital degeneracy in the frontier molecular states associated with the planar SnPc down configuration.

Importantly, this manipulation of the central metal atom operates not only at the monomer level but also extends robustly to both homodimers and heterodimers. For the heterodimer system, we use zinc phthalocyanine (ZnPc), a molecule with a higher energy gap. By exercising atomic-scale control over this structural parameter, we realize three distinct configurations in the SnPc–SnPc dimer system and two configurations in the ZnPc–SnPc dimer system. This controllability enables tunable dipole–dipole coupling and tunable energy transfer pathways, establishing an actively reconfigurable framework for engineering excitonic interactions at the molecular level.

\textbf{Controlable and reversible single-molecule optical switching}

\begin{figure}[h!]
    \centering
    \includegraphics[width=.95\textwidth]{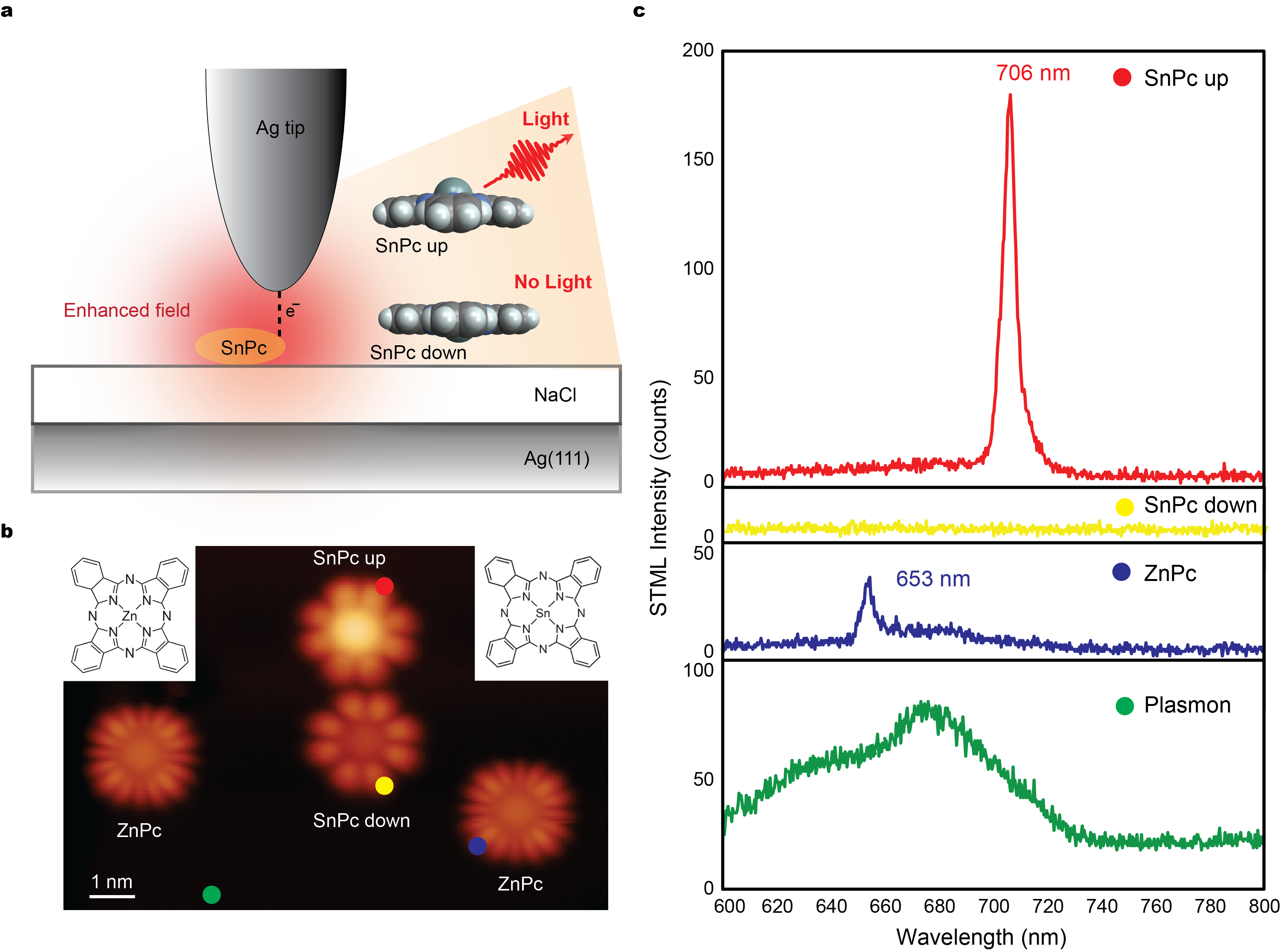}
	\caption{\textbf{Single-molecule electroluminescence through plasmon enhancement.} \textbf{a}, Schematic of STML experiment from different configurations of SnPc molecules. \textbf{b}, STM image depicting isolated single SnPc and ZnPc molecules adsorbed on 2M NaCl    island on bare Ag(111) (image size: 12 $\times$ 8 nm$^2$; scanning parameters:$V = -2.3$ V and $I$ = 6 pA). Insets: showing molecular structure of ZnPc and SnPc. \textbf{c}, STML spectra acquired at the positions marked with the ‘colored dots’ in B ($V = -2.3$ V, $I$ = 20 pA, $t = 30$ s for single-molecules and $t = 120$ s for plasmon).}
    \label{fig:sketch}
\end{figure}

Figure~\ref{fig:sketch}a schematically illustrates the experiment, in which highly localized tunneling electrons inside the STM junction are used to excite the fluorescence from individual molecule decoupled from Ag(111) by  NaCl (see section S1 in the supplementary materials). 
Figure~\ref{fig:sketch}b presents STM topographs of isolated SnPc and ZnPc molecules on 2 ML of NaCl  \cite{ploigt2007local} (see section S2 in the supplementary materials). SnPc exhibits two configurations: SnPc up (bright center, Sn atom protruding out) and SnPc down (Sn atom is planar). STML from SnPc up shows strong emission at 706 nm (1.75 eV), whereas  SnPc down exhibits negligible luminescence under identical tunneling conditions as shown in Fig.~\ref{fig:sketch}c. Photoluminescence (PL) measurements of SnPc molecules in solution show that the 706 nm transition is associated with the Q-band transition (see section S3 in the supplementary materials). Comparison of relative STML intensities reveals that SnPc up (red curve) is substantially more emissive than ZnPc (blue curve) under identical conditions, suggesting bright light molecular source as shown in Fig.~\ref{fig:sketch}c . Although the nanocavity plasmon (green curve) overlaps with both emission peaks, this overlap is necessary to enhance the molecular radiative decay rate.

These observations reveal a striking structure–property relationship: the vertical displacement of the central Sn atom appears to modulate the molecular geometry in a way that enhances or suppresses the transition dipole responsible for light emission. Such atomic-scale structural control within a single-molecule could offer a means to tune dipole–plasmon interactions and create new opportunities for engineering controllable dipole–dipole coupling and molecular-scale energy transfer processes.

\begin{figure}[t!]
    \centering
    \includegraphics[width=.95\textwidth]{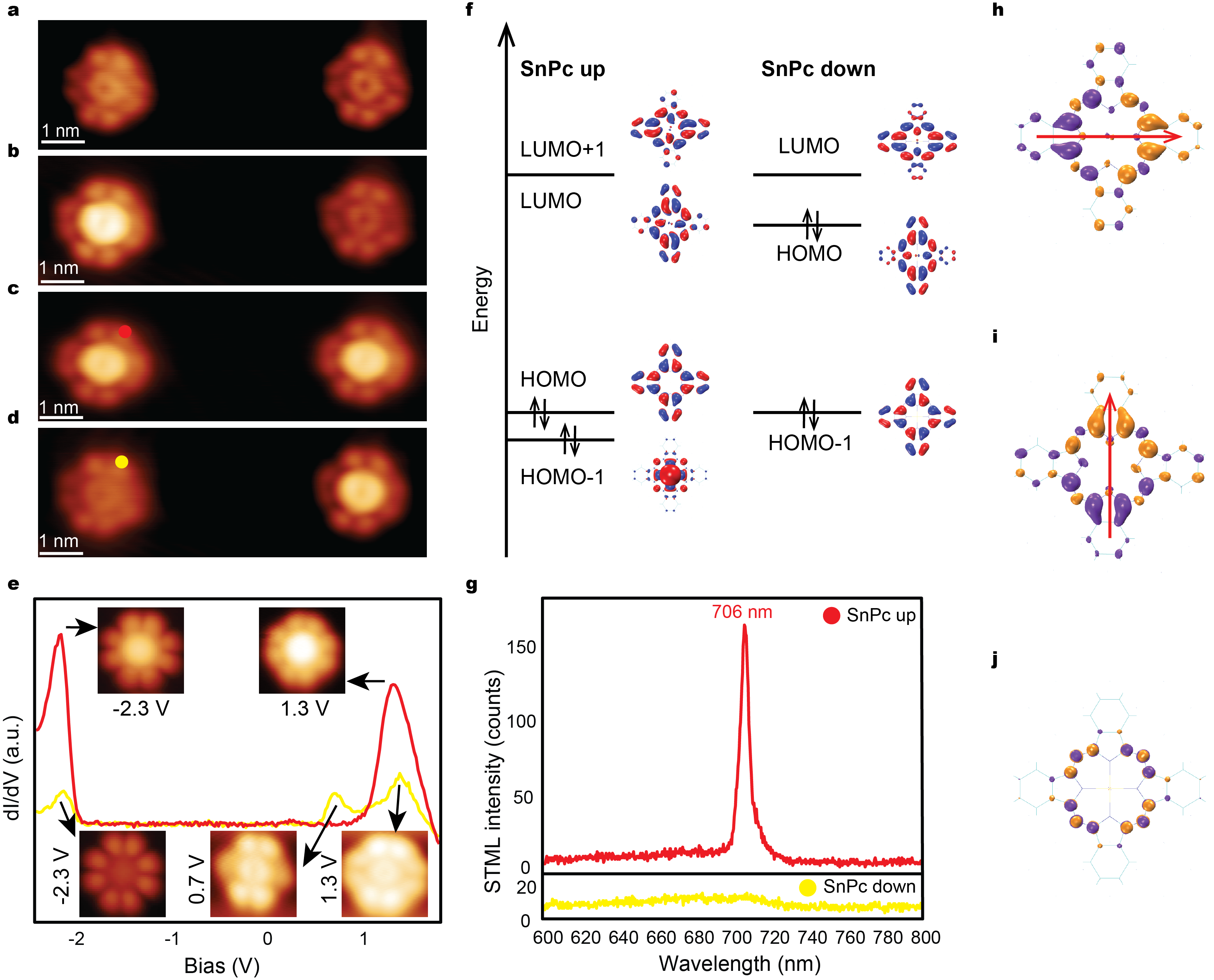}
    
	\caption{\textbf{Enhancing and suppressing the single-molecule light emission by the dipole engineering.} \textbf{a-d}, STM images depicting two isolated single SnPc molecules on 2M NaCl (image size: 9 $\times$ 3 nm$^2$ ; scanning parameters:$V = 1$ V and $I$ = 6 pA). \textbf{e}, Bias spectroscopy of SnPc up and SnPc down acquired at the positions marked with the ‘colored dots’ in c, d. \textbf{f}, Schematic energy level diagram of electron redistribution in SnPc up to SnPc down, and the DFT calculated orbitals. \textbf{g}, STML spectra of SnPc up and SnPc down acquired at the positions marked with the ‘colored dots’ in c, d ($V = -2.3$ V, $I$ = 20 pA, and $t = 30$ s). \textbf{h, i}, Calculated HOMO→LUMO+1 and HOMO→LUMO transition charge densities in the SnPc up configuration. \textbf{j}, Calculated HOMO→LUMO transition charge density in the SnPc down configuration.}
    \label{fig:switch}
\end{figure}

To identify the origin of the pronounced emission contrast between the two molecular configurations, we next examine how atomic displacement modifies the electronic structure and optical selection rules of the molecule, we demonstrate controlled and reversible switching between the two SnPc configurations (see section S4 in the supplementary materials). This is achieved by manipulating the vertical position of the central Sn atom within SnPc molecule \cite{Banerjee2025-ze, Wang2009-bo}. Briefly, both molecules were in the SnPc down configuration as shown in Fig.~\ref{fig:switch}a. A positive voltage pulse (3 V) applied to the left SnPc induced an upward displacement of the central atom, converting the molecule into the SnPc up configuration (Fig.~\ref{fig:switch}b). The same procedure was subsequently applied to the right SnPc, resulting in an identical transition  (Fig.~\ref{fig:switch}c). Reversibility was confirmed by applying a voltage pulse of opposite polarity (-3 V), which restored the left SnPc to the SnPc down state (Fig.~\ref{fig:switch}d).

Figure \ref{fig:switch}e shows a representative bias spectroscopy ($dI/dV$ ) spectra acquired on SnPc up and SnPc down configurations of same SnPc molecule. In the SnPc up configuration, the $dI/dV$ spectrum reveals two nearly symmetric resonances below ($V \sim -2.3$ V) and above ($V \sim +1.3$ V) the Fermi level. The corresponding STM images (insets in Fig. \ref{fig:switch}e) exhibit spatial patterns that closely resemble the highest occupied molecular orbital (HOMO) and the lowest unoccupied molecular orbital (LUMO) calculated using density functional theory (DFT; see Fig.~\ref{fig:switch}f and section S5 in the supplementary materials). In contrast, the SnPc down configuration reveals an additional resonance at positive bias ($V \sim +0.7$ V), alongside the two resonances observed at nearly the same energies as in the SnPc up configuration. 

DFT calculations show that in the SnPc up configuration, the LUMO is doubly degenerate, consisting of two nearly identical $\pi^*$ orbitals that remain unoccupied (Fig.~\ref{fig:switch}f). In the SnPc down configuration, this degeneracy is lifted because the Sn $5s$ lone pair hybridizes with the phthalocyanine $\pi^*$ manifold, causing partial orbital localization and reordering of the frontier levels without changing the molecular charge state (Fig.~\ref{fig:switch}f; see section S5 in the supplementary materials). One of the former LUMO components is stabilized and shifts to lower energy to become the new HOMO, while the other remains the true LUMO. Thus, the apparent shift of the “original HOMO” above the Fermi level is a consequence of degeneracy lifting and rehybridization \cite{Reecht2019-qn,Neel2016-gm} (see section S5 in the supplementary materials). 

As shown again in Fig.~\ref{fig:switch}g, the SnPc up configuration shows strong light emission, while SnPc down is non-emissive (see section S6 in the supplementary materials). Time dependent density functional theory (TDDFT) explains this contrast. In the SnPc up configuration, the Sn atom protrudes from the Pc plane, reducing the symmetry to $C_{4v}$ and relaxing selection rules. This makes the HOMO→LUMO transition optically allowed and produces a finite, mainly in-plane transition dipole (red arrows) due to asymmetry from the Sn $5s$ lone pair (Fig.~\ref{fig:switch}, h and i). The resulting nonzero dipole enables efficient plasmon–exciton coupling, leading to bright STML emission (see section S7 in the supplementary materials).
In the SnPc down configuration, the Sn atom lies nearly in-plane, restoring $D_{4h}$ symmetry. Here the HOMO→LUMO transition is dipole-forbidden because the in-plane components  related to fourfold rotational symmetry cancel, and TDDFT gives zero oscillator strength (Figure \ref{fig:switch}j), thus resulting in essentially no detectable light emission (see section S7 in the supplementary materials). These findings mark the realization of a structural parameter that controls the adsorption configuration of a single-molecule, which in turn determines its transition dipole moment and dictates its optical properties.

\begin{figure}[t!]
    \centering
    \includegraphics[width=.95\textwidth]{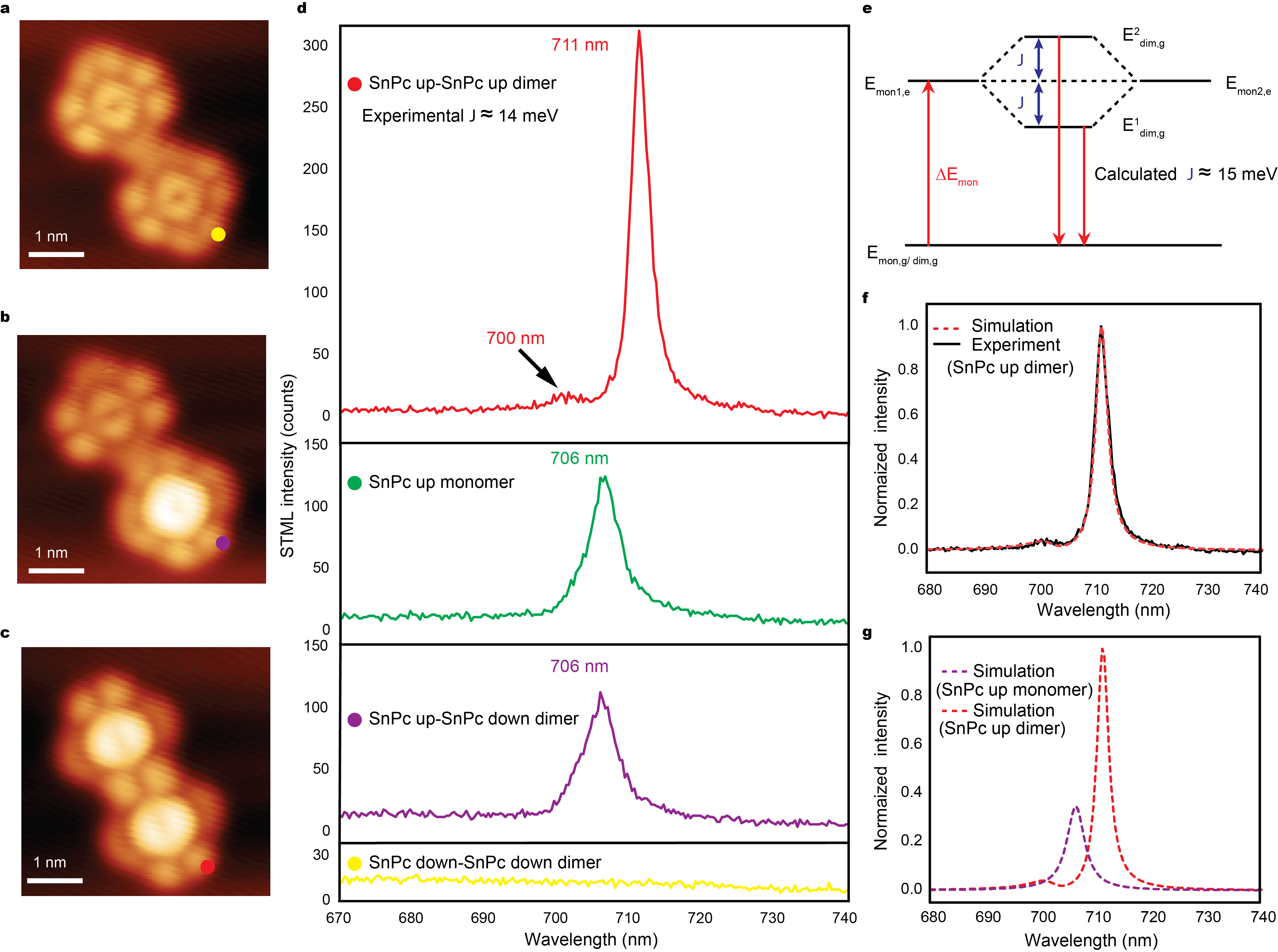}
	\caption{\textbf{Tunable coherent intermolecular dipole-dipole coupling SnPc-SnPc dimer system.} STM images revealing a SnPc dimer in three configurations on 2 ML NaCl (image size: 4.5 $\times$ 4.5 nm$^2$; scanning parameters:$V = 1$ V and $I$ = 6 pA). \textbf{a}, SnPc down-SnPc down. \textbf{b}, SnPc down-SnPc up. \textbf{c}, SnPc up-SnPc up. \textbf{d}, STML spectra of SnPc dimer acquired at the positions marked with the ‘colored dots’ with conditions ($V = -2.3$ V, $I$ = 20 pA, and $t = 30$ s) SnPc up monomer light emission (Green) is also shown for comparison. \textbf{e},  Exciton band energy diagram of a molecular dimer. $E_{\text{mon1},e}$ and $E_{\text{mon2},e}$ excited-state energy of two isolated monomer; $E_{\text{mon},g}$ ground-state energy of an of an isolated monomer;$\Delta  E_{\text{mon}}$ optical transition energy for an isolated monomer; $E_{\text{dim},g}$ ground-state energy of the dimer; $E_{\text{dim},g}^{1}$ and $E_{\text{dim},g}^{2}$ low-lying and high-lying excited-state energy of the dimer for a coupling modes with opposite phases; J exciton coupling strength. For simplicity we consider $\epsilon_{\mathrm{ex}}$ and $V$ are the same for both SnPc up monomers. \textbf{f}, Comparison between the measured and calculated STML spectrum of SnPc up-SnPc up dimer. Where the  parameters
are $\epsilon_{\mathrm{ex}} = 1.757~\mathrm{eV}$, $\epsilon_{p} = 1.9~\mathrm{eV}$, $V = 32.5~\mathrm{meV}$, and $J = 13.5~\mathrm{meV}$. \textbf{g}, Comparison between the calculated STML spectra of SnPc up monomer and SnPc up-SnPc up dimer.}
    \label{fig:homodimers}
\end{figure}

\textbf{Tunable coherent intermolecular dipole-dipole coupling in homodimer}

Having established the dipole moments for the molecular configurations, we next examined how these dipole moments mediate exciton–exciton interactions in dimers  (see also section S8 in the supplementary materials). Since a molecule with a negligible dipole moment does not contribute to light emission, it is also unlikely to participate in excitonic coupling within a dimer. Thus, a SnPc down–SnPc down dimer should remain dark, while a SnPc down–SnPc up dimer is expected to emit light similar to a single monomer. In contrast, the SnPc up–SnPc up configuration is anticipated to support multiple emission modes arising from intermolecular dipole–dipole coupling  \cite{Zhang2016-yang}.

To test this hypothesis, we designed three configurations in a dimer system—SnPc down–SnPc down, SnPc up–SnPc down, and SnPc up–SnPc up (Fig.~\ref{fig:homodimers},a, b and c)—using the same structure parameter control approach described earlier. In the SnPc down–SnPc down dimer on 2 ML NaCl (Fig.~\ref{fig:homodimers}a), neither molecule possesses a transition dipole moment under tunneling excitation, resulting in no emission and no dipole–dipole coupling (yellow curve, Fig.~\ref{fig:homodimers}d). In the SnPc down–SnPc up dimer (Fig.~\ref{fig:homodimers}b), only the SnPc-up molecule exhibits a transition dipole moment, while the SnPc-down remains inactive. Excitation on the SnPc-up lobe yields emission at 706 nm, identical to that of an isolated SnPc-up monomer (purple curve, Fig.~\ref{fig:homodimers}d). For comparison, the STML spectrum of a single SnPc-up molecule recorded under identical conditions is also shown (green curve, Fig.~\ref{fig:homodimers}d). Finally, in the SnPc up–SnPc up dimer (Fig.~\ref{fig:homodimers}c), both monomers carry transition dipole moments. The appearance of two modes at 700 nm (1.77 eV) and a high intensity mode at 711 nm (1.74 eV), together with their spectral shift from the monomer emission at 706 nm (red curve, Fig. \ref{fig:homodimers}d), indicates the presence of dipole–dipole coupling \cite{Zhang2016-yang,Doppagne2017-cq}. 

To gain insights about the excitonic coupling, which governs the interaction of adjacent molecular transition dipoles and the resulting energy splitting and optical response of the dimer, we employed a quantum mechanical modeling approach \cite{Miwa2013,Imada2017}. The system consists of two excitons and a plasmon, described by the Hamiltonian
\begin{equation}
H_{\mathrm{dimer}} = \sum_{\alpha=1}^{2} \left[ \epsilon_{\mathrm{ex}} d_{n}^{\dagger} d_{n} + \epsilon_{p} a^{\dagger} a + V (a d_{n}^{\dagger} + a^{\dagger} d_{n}) \right] 
+ J (d_{1}^{\dagger} d_{2} + d_{1} d_{2}^{\dagger}),
\end{equation}

Here, \(a^\dagger\) and \(a\) denote the plasmon creation and annihilation operators, while \(d_n^\dagger\) and \(d_n\) represent the exciton creation and annihilation operators for molecule \(n\). The parameters \(\epsilon_{\mathrm{ex}}\) and \(\epsilon_{p}\) are the exciton and plasmon energies, respectively. The quantities \(V\) and \(J\) correspond to the exciton--plasmon and exciton--exciton coupling strengths. The theoretical model yields \(J \approx 15\,\text{meV}\) (Fig.~\ref{fig:homodimers}e), which agrees well with the experimentally measured value of approximately \(14\,\text{meV}\) (Fig.~\ref{fig:homodimers}d).
This agreement clearly evidences dipole–dipole coupling, manifested in the emergence of subradiant (700 nm) and superradiant (711 nm) modes. The model quantitatively reproduces the experimental STML spectra of the SnPc up-SnPc up dimer, capturing both the peak positions and relative intensities (Fig.~\ref{fig:homodimers}f and see section S9 in the supplementary materials). To further validate the enhanced emission relative to the SnPc up monomer, we compare the normalized spectra of the dimer and monomer (Fig.~\ref{fig:homodimers}g). The calculated intensity ratio between the monomer peak at 706\,nm and the dimer peak at 711\,nm shows excellent agreement with experiment (2.87 vs.\ 2.82), underscoring the accuracy of the model in capturing the combined effects of exciton-exciton and exciton-plasmon  couplings.
 
\textbf{Controlling energy transfer in hetrodimer}

Encouraged by the demonstrated dipole control in a homodimer, we extended our structural parameter control approach to a heterogeneous ZnPc–SnPc dimer to gain insights into energy transfer process (see also section S8 in the supplementary materials).
Unlike other approaches, where energy transfer is influenced by intermolecular distance \cite{Imada2016-va} or dipole alignment \cite{Cao2021-ts} , our strategy controls energy transfer exclusively by controling the transition dipole moment of the acceptor molecule.

\begin{figure}[t!]
    \centering
    \includegraphics[width=.95\textwidth]{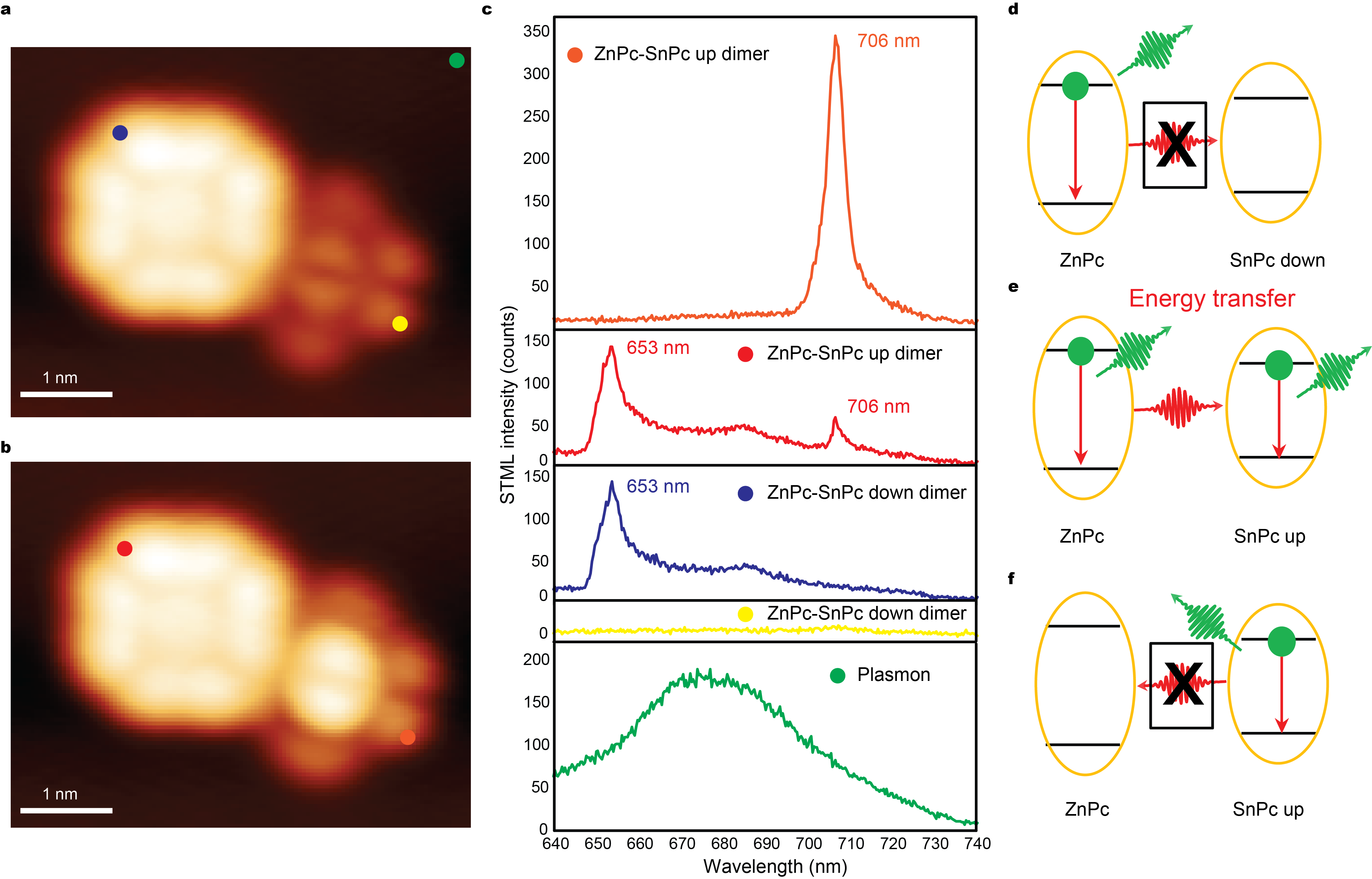}
	\caption{\textbf{Tunable Energy transfer between donor ZnPc and acceptor SnPc molecules in real space.} STM images revealing a ZnPc-SnPc dimer in two configurations on 2 ML NaCl (image size: 5 $\times$ 4 nm$^2$; scanning parameters:$V = 1$ V and $I$ = 6 pA). \textbf{a}, ZnPc-SnPc down dimer.  \textbf{b}, ZnPc-SnPc up dimer. \textbf{c}, STML spectra of SnPc dimer acquired at the positions marked with the ‘colored dots’ with conditions ($V = -2.3$ V, $I = 20$ pA, $t = 60$ s for SnPc excitation and $V = -2.5$ V, $I = 100$ pA, $t = 30$ s for ZnPc and plasmon excitation ). \textbf{d}, Schematics of showing no energy transfer from ZnPc-SnPc down dimer configuration, if excitation is from
ZnPc side. \textbf{e}, Schematics of showing energy transfer from ZnPc-SnPc up dimer configuration, if excitation is from
ZnPc side. \textbf{f}, Schematics of showing no energy transfer from ZnPc-SnPc up dimer configuration, if excitation is from SnPc side.}
    \label{fig:heterodimers}
\end{figure}
Two distinct configurations were designed, namely ZnPc–SnPc down and ZnPc–SnPc up (Fig.~\ref{fig:heterodimers}, a and b). In the ZnPc–SnPc down configuration, the STML spectrum acquired at the SnPc (yellow dot in Fig.~\ref{fig:heterodimers}a; yellow curve in Fig.~\ref{fig:heterodimers}c) showed no detectable emission, as expected, since this configuration does not support a dipole moment. Conversely, excitation at the ZnPc site (blue dot in Fig.~\ref{fig:heterodimers}a; blue curve in Fig.~\ref{fig:heterodimers}c) revealed a characteristic ZnPc electroluminescence spectrum. These results indicate that energy transfer from the donor (ZnPc) to the acceptor (SnPc down) does not occur when the acceptor molecule has a negligible dipole moment, as illustrated schematically in Fig.~\ref{fig:heterodimers}d.

In contrast, the ZnPc–SnPc up configuration exhibited markedly different behavior. Excitation at the ZnPc site (red dot in Fig.~\ref{fig:heterodimers}b; red curve in Fig.~\ref{fig:heterodimers}c) produced emission from both ZnPc and SnPc up, demonstrating efficient donor–acceptor energy transfer, facilitated by the finite dipole moment of SnPc up in this configuration (schematic in Fig.~\ref{fig:heterodimers}e). This finding suggests that the presence of dipoles in both the donor and the acceptor is required for energy transfer, indicating resonant energy transfer. Furthermore, excitation at the SnPc site (orange dot in Fig.~\ref{fig:heterodimers}b; orange curve in Fig.~\ref{fig:heterodimers}c) yielded only the characteristic SnPc electroluminescence, as expected. However, no reverse energy transfer to ZnPc was observed, consistent with the fact that the emission energy of SnPc up is lower than that of ZnPc (schematic in Fig.~\ref{fig:heterodimers}f). We have demonstrated that resonant energy transfer within a heterogeneous ZnPc–SnPc dimer can be selectively controlled by modulating the dipole moment of the acceptor molecule.

\textbf{Conclusions}

 In conclusion, we demonstrate atomic-scale control within a single-molecule that enables deliberate reconfiguration of its transition dipole moment and on-demand switching of light emission. This capability establishes a molecular platform for tuning collective optical responses with atomic precision. Beyond its immediate implications, our approach provides new opportunities to interrogate plasmon–exciton coupling, tunable dipole–dipole interactions, and nanoscale energy-transfer dynamics, while laying the groundwork for single-molecule quantum engineering and atomically reconfigurable optoelectronic devices.

\section*{Data availability}
All data needed to evaluate the conclusions in the paper are present in the paper or the supplementary materials. Data for all figures presented in this study are available on Zenodo. The following software was used: ORCA to perform ab initio calculations on the molecules, VMD  for visualization, and Matlab R2021a and Multiwfn for data postprocessing. Gwyddion to visualize the STM topography images. Python for the effective model analysis and plotting.

\section*{Acknowledgements}
This research made use of Nanoscience Center (NSC) at the University of  Jyväskyl\"a (JYU) facilities and was supported by the European Research Council (ERC-2021-StG
No. 101039500 “Tailoring Quantum Matter on the
Flatland”), Research Council of Finland (Academy Project nos.~3370910), Jane and Aatos Erkko Foundation (Project EffQSim), The Finnish Ministry of Education and Culture through the Quantum Doctoral Education Pilot Program (QDOC VN/3137/2024-OKM-4),
InstituteQ,
the ERC Consolidator Grant ULTRATWISTROICS (Grant agreement no. 101170477),
and the Research Council of Finland through the Finnish Quantum Flagship project (Project Nos. 359240 and 358877). We acknowledge Finnish Centre of Excellence in Quantum Materials (QMAT) and CSC–IT Center for Science, Finland, for computational resources and the Aalto Science-IT project. 

\section*{Author contributions}
AG, SN, and SK conceived the experiment. YS and JL developed the theoretical effective model. AG, SN, TT  carried out the low-temperature STM experiments. AG analyzed the experimental data. OMS, RT, and YP carried out the DFT calculations. SK and AG wrote the manuscript with input from all coauthors. All authors contributed to editing and preparing the manuscript.
\section*{Competing interests}
The authors declare no competing interests.

\bibliography{ref}

\begin{thebibliography}{35}%
\makeatletter
\providecommand \@ifxundefined [1]{%
 \@ifx{#1\undefined}
}%
\providecommand \@ifnum [1]{%
 \ifnum #1\expandafter \@firstoftwo
 \else \expandafter \@secondoftwo
 \fi
}%
\providecommand \@ifx [1]{%
 \ifx #1\expandafter \@firstoftwo
 \else \expandafter \@secondoftwo
 \fi
}%
\providecommand \natexlab [1]{#1}%
\providecommand \enquote  [1]{``#1''}%
\providecommand \bibnamefont  [1]{#1}%
\providecommand \bibfnamefont [1]{#1}%
\providecommand \citenamefont [1]{#1}%
\providecommand \href@noop [0]{\@secondoftwo}%
\providecommand \href [0]{\begingroup \@sanitize@url \@href}%
\providecommand \@href[1]{\@@startlink{#1}\@@href}%
\providecommand \@@href[1]{\endgroup#1\@@endlink}%
\providecommand \@sanitize@url [0]{\catcode `\\12\catcode `\$12\catcode `\&12\catcode `\#12\catcode `\^12\catcode `\_12\catcode `\%12\relax}%
\providecommand \@@startlink[1]{}%
\providecommand \@@endlink[0]{}%
\providecommand \url  [0]{\begingroup\@sanitize@url \@url }%
\providecommand \@url [1]{\endgroup\@href {#1}{\urlprefix }}%
\providecommand \urlprefix  [0]{URL }%
\providecommand \Eprint [0]{\href }%
\providecommand \doibase [0]{http://dx.doi.org/}%
\providecommand \selectlanguage [0]{\@gobble}%
\providecommand \bibinfo  [0]{\@secondoftwo}%
\providecommand \bibfield  [0]{\@secondoftwo}%
\providecommand \translation [1]{[#1]}%
\providecommand \BibitemOpen [0]{}%
\providecommand \bibitemStop [0]{}%
\providecommand \bibitemNoStop [0]{.\EOS\space}%
\providecommand \EOS [0]{\spacefactor3000\relax}%
\providecommand \BibitemShut  [1]{\csname bibitem#1\endcsname}%
\let\auto@bib@innerbib\@empty
\bibitem [{\citenamefont {Feringa}\ and\ \citenamefont {Browne}(2011)}]{feringa2011molecular}%
  \BibitemOpen
  \bibfield  {author} {\bibinfo {author} {\bibfnamefont {Ben~L}\ \bibnamefont {Feringa}}\ and\ \bibinfo {author} {\bibfnamefont {Wesley~R}\ \bibnamefont {Browne}},\ }\href@noop {} {\emph {\bibinfo {title} {Molecular switches}}}\ (\bibinfo  {publisher} {John Wiley \& Sons},\ \bibinfo {year} {2011})\BibitemShut {NoStop}%
\bibitem [{\citenamefont {Vogelsang}\ \emph {et~al.}(2009)\citenamefont {Vogelsang}, \citenamefont {Cordes}, \citenamefont {Forthmann}, \citenamefont {Steinhauer},\ and\ \citenamefont {Tinnefeld}}]{vogelsang2009controlling}%
  \BibitemOpen
  \bibfield  {author} {\bibinfo {author} {\bibfnamefont {Jan}\ \bibnamefont {Vogelsang}}, \bibinfo {author} {\bibfnamefont {Thorben}\ \bibnamefont {Cordes}}, \bibinfo {author} {\bibfnamefont {Carsten}\ \bibnamefont {Forthmann}}, \bibinfo {author} {\bibfnamefont {Christian}\ \bibnamefont {Steinhauer}}, \ and\ \bibinfo {author} {\bibfnamefont {Philip}\ \bibnamefont {Tinnefeld}},\ }\bibfield  {title} {\enquote {\bibinfo {title} {Controlling the fluorescence of ordinary oxazine dyes for single-molecule switching and superresolution microscopy},}\ }\href@noop {} {\bibfield  {journal} {\bibinfo  {journal} {Proceedings of the National Academy of Sciences}\ }\textbf {\bibinfo {volume} {106}},\ \bibinfo {pages} {8107--8112} (\bibinfo {year} {2009})}\BibitemShut {NoStop}%
\bibitem [{\citenamefont {Basch{\'e}}\ \emph {et~al.}(1995)\citenamefont {Basch{\'e}}, \citenamefont {Kummer},\ and\ \citenamefont {Br{\"a}uchle}}]{basche1995direct}%
  \BibitemOpen
  \bibfield  {author} {\bibinfo {author} {\bibfnamefont {Th}~\bibnamefont {Basch{\'e}}}, \bibinfo {author} {\bibfnamefont {S}~\bibnamefont {Kummer}}, \ and\ \bibinfo {author} {\bibfnamefont {Ch}~\bibnamefont {Br{\"a}uchle}},\ }\bibfield  {title} {\enquote {\bibinfo {title} {Direct spectroscopic observation of quantum jumps of a single molecule},}\ }\href@noop {} {\bibfield  {journal} {\bibinfo  {journal} {Nature}\ }\textbf {\bibinfo {volume} {373}},\ \bibinfo {pages} {132--134} (\bibinfo {year} {1995})}\BibitemShut {NoStop}%
\bibitem [{\citenamefont {Zondervan}\ \emph {et~al.}(2003)\citenamefont {Zondervan}, \citenamefont {Kulzer}, \citenamefont {Orlinskii},\ and\ \citenamefont {Orrit}}]{zondervan2003photoblinking}%
  \BibitemOpen
  \bibfield  {author} {\bibinfo {author} {\bibfnamefont {Rob}\ \bibnamefont {Zondervan}}, \bibinfo {author} {\bibfnamefont {Florian}\ \bibnamefont {Kulzer}}, \bibinfo {author} {\bibfnamefont {Sergei~B}\ \bibnamefont {Orlinskii}}, \ and\ \bibinfo {author} {\bibfnamefont {Michel}\ \bibnamefont {Orrit}},\ }\bibfield  {title} {\enquote {\bibinfo {title} {Photoblinking of rhodamine 6g in poly (vinyl alcohol): Radical dark state formed through the triplet},}\ }\href@noop {} {\bibfield  {journal} {\bibinfo  {journal} {The Journal of Physical Chemistry A}\ }\textbf {\bibinfo {volume} {107}},\ \bibinfo {pages} {6770--6776} (\bibinfo {year} {2003})}\BibitemShut {NoStop}%
\bibitem [{\citenamefont {Kaiser}\ \emph {et~al.}(2024)\citenamefont {Kaiser}, \citenamefont {Jiang}, \citenamefont {Romeo}, \citenamefont {Scheurer}, \citenamefont {Schull},\ and\ \citenamefont {Ros{\l}awska}}]{kaiser2024gating}%
  \BibitemOpen
  \bibfield  {author} {\bibinfo {author} {\bibfnamefont {Katharina}\ \bibnamefont {Kaiser}}, \bibinfo {author} {\bibfnamefont {Song}\ \bibnamefont {Jiang}}, \bibinfo {author} {\bibfnamefont {Michelangelo}\ \bibnamefont {Romeo}}, \bibinfo {author} {\bibfnamefont {Fabrice}\ \bibnamefont {Scheurer}}, \bibinfo {author} {\bibfnamefont {Guillaume}\ \bibnamefont {Schull}}, \ and\ \bibinfo {author} {\bibfnamefont {Anna}\ \bibnamefont {Ros{\l}awska}},\ }\bibfield  {title} {\enquote {\bibinfo {title} {Gating single-molecule fluorescence with electrons},}\ }\href@noop {} {\bibfield  {journal} {\bibinfo  {journal} {Physical Review Letters}\ }\textbf {\bibinfo {volume} {133}},\ \bibinfo {pages} {156902} (\bibinfo {year} {2024})}\BibitemShut {NoStop}%
\bibitem [{\citenamefont {Rasnik}\ \emph {et~al.}(2006)\citenamefont {Rasnik}, \citenamefont {McKinney},\ and\ \citenamefont {Ha}}]{rasnik2006nonblinking}%
  \BibitemOpen
  \bibfield  {author} {\bibinfo {author} {\bibfnamefont {Ivan}\ \bibnamefont {Rasnik}}, \bibinfo {author} {\bibfnamefont {Sean~A}\ \bibnamefont {McKinney}}, \ and\ \bibinfo {author} {\bibfnamefont {Taekjip}\ \bibnamefont {Ha}},\ }\bibfield  {title} {\enquote {\bibinfo {title} {Nonblinking and long-lasting single-molecule fluorescence imaging},}\ }\href@noop {} {\bibfield  {journal} {\bibinfo  {journal} {Nature methods}\ }\textbf {\bibinfo {volume} {3}},\ \bibinfo {pages} {891--893} (\bibinfo {year} {2006})}\BibitemShut {NoStop}%
\bibitem [{\citenamefont {Vogelsang}\ \emph {et~al.}(2008)\citenamefont {Vogelsang}, \citenamefont {Kasper}, \citenamefont {Steinhauer}, \citenamefont {Person}, \citenamefont {Heilemann}, \citenamefont {Sauer},\ and\ \citenamefont {Tinnefeld}}]{vogelsang2008reducing}%
  \BibitemOpen
  \bibfield  {author} {\bibinfo {author} {\bibfnamefont {Jan}\ \bibnamefont {Vogelsang}}, \bibinfo {author} {\bibfnamefont {Robert}\ \bibnamefont {Kasper}}, \bibinfo {author} {\bibfnamefont {Christian}\ \bibnamefont {Steinhauer}}, \bibinfo {author} {\bibfnamefont {Britta}\ \bibnamefont {Person}}, \bibinfo {author} {\bibfnamefont {Mike}\ \bibnamefont {Heilemann}}, \bibinfo {author} {\bibfnamefont {Markus}\ \bibnamefont {Sauer}}, \ and\ \bibinfo {author} {\bibfnamefont {Philip}\ \bibnamefont {Tinnefeld}},\ }\bibfield  {title} {\enquote {\bibinfo {title} {A reducing and oxidizing system minimizes photobleaching and blinking of fluorescent dyes},}\ }\href@noop {} {\bibfield  {journal} {\bibinfo  {journal} {Angewandte Chemie-International Edition}\ }\textbf {\bibinfo {volume} {47}} (\bibinfo {year} {2008})}\BibitemShut {NoStop}%
\bibitem [{\citenamefont {Kuhnke}\ \emph {et~al.}(2017)\citenamefont {Kuhnke}, \citenamefont {Gro{\ss}e}, \citenamefont {Merino},\ and\ \citenamefont {Kern}}]{kuhnke2017atomic}%
  \BibitemOpen
  \bibfield  {author} {\bibinfo {author} {\bibfnamefont {Klaus}\ \bibnamefont {Kuhnke}}, \bibinfo {author} {\bibfnamefont {Christoph}\ \bibnamefont {Gro{\ss}e}}, \bibinfo {author} {\bibfnamefont {Pablo}\ \bibnamefont {Merino}}, \ and\ \bibinfo {author} {\bibfnamefont {Klaus}\ \bibnamefont {Kern}},\ }\bibfield  {title} {\enquote {\bibinfo {title} {Atomic-scale imaging and spectroscopy of electroluminescence at molecular interfaces},}\ }\href@noop {} {\bibfield  {journal} {\bibinfo  {journal} {Chemical reviews}\ }\textbf {\bibinfo {volume} {117}},\ \bibinfo {pages} {5174--5222} (\bibinfo {year} {2017})}\BibitemShut {NoStop}%
\bibitem [{\citenamefont {Yang}\ \emph {et~al.}(2020)\citenamefont {Yang}, \citenamefont {Chen}, \citenamefont {Ghafoor}, \citenamefont {Zhang}, \citenamefont {Zhang}, \citenamefont {Zhang}, \citenamefont {Luo}, \citenamefont {Yang}, \citenamefont {Sandoghdar}, \citenamefont {Aizpurua} \emph {et~al.}}]{yang2020sub}%
  \BibitemOpen
  \bibfield  {author} {\bibinfo {author} {\bibfnamefont {Ben}\ \bibnamefont {Yang}}, \bibinfo {author} {\bibfnamefont {Gong}\ \bibnamefont {Chen}}, \bibinfo {author} {\bibfnamefont {Atif}\ \bibnamefont {Ghafoor}}, \bibinfo {author} {\bibfnamefont {Yufan}\ \bibnamefont {Zhang}}, \bibinfo {author} {\bibfnamefont {Yao}\ \bibnamefont {Zhang}}, \bibinfo {author} {\bibfnamefont {Yang}\ \bibnamefont {Zhang}}, \bibinfo {author} {\bibfnamefont {Yi}~\bibnamefont {Luo}}, \bibinfo {author} {\bibfnamefont {Jinlong}\ \bibnamefont {Yang}}, \bibinfo {author} {\bibfnamefont {Vahid}\ \bibnamefont {Sandoghdar}}, \bibinfo {author} {\bibfnamefont {Javier}\ \bibnamefont {Aizpurua}},  \emph {et~al.},\ }\bibfield  {title} {\enquote {\bibinfo {title} {Sub-nanometre resolution in single-molecule photoluminescence imaging},}\ }\href@noop {} {\bibfield  {journal} {\bibinfo  {journal} {Nature Photonics}\ }\textbf {\bibinfo {volume} {14}},\ \bibinfo {pages} {693--699} (\bibinfo {year} {2020})}\BibitemShut {NoStop}%
\bibitem [{\citenamefont {Lee}\ \emph {et~al.}(2014)\citenamefont {Lee}, \citenamefont {Perdue}, \citenamefont {Rodriguez~Perez},\ and\ \citenamefont {Apkarian}}]{lee2014vibronic}%
  \BibitemOpen
  \bibfield  {author} {\bibinfo {author} {\bibfnamefont {Joonhee}\ \bibnamefont {Lee}}, \bibinfo {author} {\bibfnamefont {Shawn~M}\ \bibnamefont {Perdue}}, \bibinfo {author} {\bibfnamefont {Alejandro}\ \bibnamefont {Rodriguez~Perez}}, \ and\ \bibinfo {author} {\bibfnamefont {Vartkess~Ara}\ \bibnamefont {Apkarian}},\ }\bibfield  {title} {\enquote {\bibinfo {title} {Vibronic motion with joint angstrom--femtosecond resolution observed through fano progressions recorded within one molecule},}\ }\href@noop {} {\bibfield  {journal} {\bibinfo  {journal} {ACS nano}\ }\textbf {\bibinfo {volume} {8}},\ \bibinfo {pages} {54--63} (\bibinfo {year} {2014})}\BibitemShut {NoStop}%
\bibitem [{\citenamefont {Schneider}\ and\ \citenamefont {Berndt}(2012)}]{schneider2012plasmonic}%
  \BibitemOpen
  \bibfield  {author} {\bibinfo {author} {\bibfnamefont {Natalia~L}\ \bibnamefont {Schneider}}\ and\ \bibinfo {author} {\bibfnamefont {Richard}\ \bibnamefont {Berndt}},\ }\bibfield  {title} {\enquote {\bibinfo {title} {Plasmonic excitation of light emission and absorption by porphyrine molecules in a scanning tunneling microscope},}\ }\href@noop {} {\bibfield  {journal} {\bibinfo  {journal} {Physical Review B—Condensed Matter and Materials Physics}\ }\textbf {\bibinfo {volume} {86}},\ \bibinfo {pages} {035445} (\bibinfo {year} {2012})}\BibitemShut {NoStop}%
\bibitem [{\citenamefont {Qiu}\ \emph {et~al.}(2003)\citenamefont {Qiu}, \citenamefont {Nazin},\ and\ \citenamefont {Ho}}]{qiu2003vibrationally}%
  \BibitemOpen
  \bibfield  {author} {\bibinfo {author} {\bibfnamefont {XH}~\bibnamefont {Qiu}}, \bibinfo {author} {\bibfnamefont {GV}~\bibnamefont {Nazin}}, \ and\ \bibinfo {author} {\bibfnamefont {W}~\bibnamefont {Ho}},\ }\bibfield  {title} {\enquote {\bibinfo {title} {Vibrationally resolved fluorescence excited with submolecular precision},}\ }\href@noop {} {\bibfield  {journal} {\bibinfo  {journal} {Science}\ }\textbf {\bibinfo {volume} {299}},\ \bibinfo {pages} {542--546} (\bibinfo {year} {2003})}\BibitemShut {NoStop}%
\bibitem [{\citenamefont {Grewal}\ \emph {et~al.}(2025)\citenamefont {Grewal}, \citenamefont {Imada}, \citenamefont {Miwa}, \citenamefont {Imai-Imada}, \citenamefont {Kimura}, \citenamefont {Jaculbia}, \citenamefont {Kuhnke}, \citenamefont {Kern},\ and\ \citenamefont {Kim}}]{acsnano.5c04193}%
  \BibitemOpen
  \bibfield  {author} {\bibinfo {author} {\bibfnamefont {Abhishek}\ \bibnamefont {Grewal}}, \bibinfo {author} {\bibfnamefont {Hiroshi}\ \bibnamefont {Imada}}, \bibinfo {author} {\bibfnamefont {Kuniyuki}\ \bibnamefont {Miwa}}, \bibinfo {author} {\bibfnamefont {Miyabi}\ \bibnamefont {Imai-Imada}}, \bibinfo {author} {\bibfnamefont {Kensuke}\ \bibnamefont {Kimura}}, \bibinfo {author} {\bibfnamefont {Rafael}\ \bibnamefont {Jaculbia}}, \bibinfo {author} {\bibfnamefont {Klaus}\ \bibnamefont {Kuhnke}}, \bibinfo {author} {\bibfnamefont {Klaus}\ \bibnamefont {Kern}}, \ and\ \bibinfo {author} {\bibfnamefont {Yousoo}\ \bibnamefont {Kim}},\ }\bibfield  {title} {\enquote {\bibinfo {title} {Single-molecule phosphorescence and intersystem crossing in a coupled exciton plasmon system},}\ }\href@noop {} {\bibfield  {journal} {\bibinfo  {journal} {ACS Nano}\ }\textbf {\bibinfo {volume} {19}},\ \bibinfo {pages} {23796--23805} (\bibinfo {year} {2025})}\BibitemShut {NoStop}%
\bibitem [{\citenamefont {Zhu}\ \emph {et~al.}(2025)\citenamefont {Zhu}, \citenamefont {Zhang}, \citenamefont {Zhang}, \citenamefont {Cui}, \citenamefont {Li}, \citenamefont {Zhang}, \citenamefont {Yang}, \citenamefont {Chen}, \citenamefont {Zhang}, \citenamefont {Dong} \emph {et~al.}}]{zhu2025probing}%
  \BibitemOpen
  \bibfield  {author} {\bibinfo {author} {\bibfnamefont {Rui}\ \bibnamefont {Zhu}}, \bibinfo {author} {\bibfnamefont {Yi-Hao}\ \bibnamefont {Zhang}}, \bibinfo {author} {\bibfnamefont {Yu-Fan}\ \bibnamefont {Zhang}}, \bibinfo {author} {\bibfnamefont {Jie}\ \bibnamefont {Cui}}, \bibinfo {author} {\bibfnamefont {Hang}\ \bibnamefont {Li}}, \bibinfo {author} {\bibfnamefont {Xian-Biao}\ \bibnamefont {Zhang}}, \bibinfo {author} {\bibfnamefont {Ben}\ \bibnamefont {Yang}}, \bibinfo {author} {\bibfnamefont {Gong}\ \bibnamefont {Chen}}, \bibinfo {author} {\bibfnamefont {Yao}\ \bibnamefont {Zhang}}, \bibinfo {author} {\bibfnamefont {Zhen-Chao}\ \bibnamefont {Dong}},  \emph {et~al.},\ }\bibfield  {title} {\enquote {\bibinfo {title} {Probing vibronic coupling in molecular oligomers with 1-0 resonance tip-enhanced raman spectroscopy},}\ }\href@noop {} {\bibfield  {journal} {\bibinfo  {journal} {Chinese Journal of Chemical Physics}\ } (\bibinfo {year} {2025})}\BibitemShut {NoStop}%
\bibitem [{\citenamefont {Doppagne}\ \emph {et~al.}(2018)\citenamefont {Doppagne}, \citenamefont {Chong}, \citenamefont {Bulou}, \citenamefont {Boeglin}, \citenamefont {Scheurer},\ and\ \citenamefont {Schull}}]{Doppagne2018-science}%
  \BibitemOpen
  \bibfield  {author} {\bibinfo {author} {\bibfnamefont {Benjamin}\ \bibnamefont {Doppagne}}, \bibinfo {author} {\bibfnamefont {Michael~C.}\ \bibnamefont {Chong}}, \bibinfo {author} {\bibfnamefont {Hervé}\ \bibnamefont {Bulou}}, \bibinfo {author} {\bibfnamefont {Alex}\ \bibnamefont {Boeglin}}, \bibinfo {author} {\bibfnamefont {Fabrice}\ \bibnamefont {Scheurer}}, \ and\ \bibinfo {author} {\bibfnamefont {Guillaume}\ \bibnamefont {Schull}},\ }\bibfield  {title} {\enquote {\bibinfo {title} {Electrofluorochromism at the single-molecule level},}\ }\href@noop {} {\bibfield  {journal} {\bibinfo  {journal} {Science}\ }\textbf {\bibinfo {volume} {361}},\ \bibinfo {pages} {251--255} (\bibinfo {year} {2018})}\BibitemShut {NoStop}%
\bibitem [{\citenamefont {Chen}\ \emph {et~al.}(2010)\citenamefont {Chen}, \citenamefont {Chu}, \citenamefont {Bobisch}, \citenamefont {Mills},\ and\ \citenamefont {Ho}}]{chen2010viewing}%
  \BibitemOpen
  \bibfield  {author} {\bibinfo {author} {\bibfnamefont {Chi}\ \bibnamefont {Chen}}, \bibinfo {author} {\bibfnamefont {Ping}\ \bibnamefont {Chu}}, \bibinfo {author} {\bibfnamefont {CA}~\bibnamefont {Bobisch}}, \bibinfo {author} {\bibfnamefont {DL}~\bibnamefont {Mills}}, \ and\ \bibinfo {author} {\bibfnamefont {W}~\bibnamefont {Ho}},\ }\bibfield  {title} {\enquote {\bibinfo {title} {Viewing the interior of a single molecule: Vibronically resolved photon imaging at submolecular resolution},}\ }\href@noop {} {\bibfield  {journal} {\bibinfo  {journal} {Physical Review Letters}\ }\textbf {\bibinfo {volume} {105}},\ \bibinfo {pages} {217402} (\bibinfo {year} {2010})}\BibitemShut {NoStop}%
\bibitem [{\citenamefont {Imada}\ \emph {et~al.}(2016)\citenamefont {Imada}, \citenamefont {Miwa}, \citenamefont {Imai-Imada}, \citenamefont {Kawahara}, \citenamefont {Kimura},\ and\ \citenamefont {Kim}}]{Imada2016-va}%
  \BibitemOpen
  \bibfield  {author} {\bibinfo {author} {\bibfnamefont {Hiroshi}\ \bibnamefont {Imada}}, \bibinfo {author} {\bibfnamefont {Kuniyuki}\ \bibnamefont {Miwa}}, \bibinfo {author} {\bibfnamefont {Miyabi}\ \bibnamefont {Imai-Imada}}, \bibinfo {author} {\bibfnamefont {Shota}\ \bibnamefont {Kawahara}}, \bibinfo {author} {\bibfnamefont {Kensuke}\ \bibnamefont {Kimura}}, \ and\ \bibinfo {author} {\bibfnamefont {Yousoo}\ \bibnamefont {Kim}},\ }\bibfield  {title} {\enquote {\bibinfo {title} {Real-space investigation of energy transfer in heterogeneous molecular dimers},}\ }\href@noop {} {\bibfield  {journal} {\bibinfo  {journal} {Nature}\ }\textbf {\bibinfo {volume} {538}},\ \bibinfo {pages} {364--367} (\bibinfo {year} {2016})}\BibitemShut {NoStop}%
\bibitem [{\citenamefont {Zhang}\ \emph {et~al.}(2016)\citenamefont {Zhang}, \citenamefont {Luo}, \citenamefont {Zhang}, \citenamefont {Yu}, \citenamefont {Kuang}, \citenamefont {Zhang}, \citenamefont {Meng}, \citenamefont {Luo}, \citenamefont {Yang}, \citenamefont {Dong},\ and\ \citenamefont {Hou}}]{Zhang2016-yang}%
  \BibitemOpen
  \bibfield  {author} {\bibinfo {author} {\bibfnamefont {Yang}\ \bibnamefont {Zhang}}, \bibinfo {author} {\bibfnamefont {Yang}\ \bibnamefont {Luo}}, \bibinfo {author} {\bibfnamefont {Yao}\ \bibnamefont {Zhang}}, \bibinfo {author} {\bibfnamefont {Yun-Jie}\ \bibnamefont {Yu}}, \bibinfo {author} {\bibfnamefont {Yan-Min}\ \bibnamefont {Kuang}}, \bibinfo {author} {\bibfnamefont {Li}~\bibnamefont {Zhang}}, \bibinfo {author} {\bibfnamefont {Qiu-Shi}\ \bibnamefont {Meng}}, \bibinfo {author} {\bibfnamefont {Yi}~\bibnamefont {Luo}}, \bibinfo {author} {\bibfnamefont {Jin-Long}\ \bibnamefont {Yang}}, \bibinfo {author} {\bibfnamefont {Zhen-Chao}\ \bibnamefont {Dong}}, \ and\ \bibinfo {author} {\bibfnamefont {J~G}\ \bibnamefont {Hou}},\ }\bibfield  {title} {\enquote {\bibinfo {title} {Visualizing coherent intermolecular dipole--dipole coupling in real space},}\ }\href@noop {} {\bibfield  {journal} {\bibinfo  {journal} {Nature}\ }\textbf {\bibinfo {volume} {531}},\ \bibinfo {pages} {623--627} (\bibinfo {year}
  {2016})}\BibitemShut {NoStop}%
\bibitem [{\citenamefont {Cao}\ \emph {et~al.}(2021)\citenamefont {Cao}, \citenamefont {Ros{\l}awska}, \citenamefont {Doppagne}, \citenamefont {Romeo}, \citenamefont {F{\'e}ron}, \citenamefont {Ch{\'e}rioux}, \citenamefont {Bulou}, \citenamefont {Scheurer},\ and\ \citenamefont {Schull}}]{Cao2021-ts}%
  \BibitemOpen
  \bibfield  {author} {\bibinfo {author} {\bibfnamefont {Shuiyan}\ \bibnamefont {Cao}}, \bibinfo {author} {\bibfnamefont {Anna}\ \bibnamefont {Ros{\l}awska}}, \bibinfo {author} {\bibfnamefont {Benjamin}\ \bibnamefont {Doppagne}}, \bibinfo {author} {\bibfnamefont {Michelangelo}\ \bibnamefont {Romeo}}, \bibinfo {author} {\bibfnamefont {Michel}\ \bibnamefont {F{\'e}ron}}, \bibinfo {author} {\bibfnamefont {Fr{\'e}d{\'e}ric}\ \bibnamefont {Ch{\'e}rioux}}, \bibinfo {author} {\bibfnamefont {Herv{\'e}}\ \bibnamefont {Bulou}}, \bibinfo {author} {\bibfnamefont {Fabrice}\ \bibnamefont {Scheurer}}, \ and\ \bibinfo {author} {\bibfnamefont {Guillaume}\ \bibnamefont {Schull}},\ }\bibfield  {title} {\enquote {\bibinfo {title} {Energy funnelling within multichromophore architectures monitored with subnanometre resolution},}\ }\href@noop {} {\bibfield  {journal} {\bibinfo  {journal} {Nature Chemistry}\ }\textbf {\bibinfo {volume} {13}},\ \bibinfo {pages} {766--770} (\bibinfo {year} {2021})}\BibitemShut {NoStop}%
\bibitem [{\citenamefont {Hung}\ \emph {et~al.}(2024)\citenamefont {Hung}, \citenamefont {Godinez-Loyola}, \citenamefont {Steinbrecher}, \citenamefont {Kiraly}, \citenamefont {Khajetoorians}, \citenamefont {Doltsinis}, \citenamefont {Strassert},\ and\ \citenamefont {Wegner}}]{hung2024activating}%
  \BibitemOpen
  \bibfield  {author} {\bibinfo {author} {\bibfnamefont {Tzu-Chao}\ \bibnamefont {Hung}}, \bibinfo {author} {\bibfnamefont {Yokari}\ \bibnamefont {Godinez-Loyola}}, \bibinfo {author} {\bibfnamefont {Manuel}\ \bibnamefont {Steinbrecher}}, \bibinfo {author} {\bibfnamefont {Brian}\ \bibnamefont {Kiraly}}, \bibinfo {author} {\bibfnamefont {Alexander~A}\ \bibnamefont {Khajetoorians}}, \bibinfo {author} {\bibfnamefont {Nikos~L}\ \bibnamefont {Doltsinis}}, \bibinfo {author} {\bibfnamefont {Cristian~A}\ \bibnamefont {Strassert}}, \ and\ \bibinfo {author} {\bibfnamefont {Daniel}\ \bibnamefont {Wegner}},\ }\bibfield  {title} {\enquote {\bibinfo {title} {Activating the fluorescence of a ni (ii) complex by energy transfer},}\ }\href@noop {} {\bibfield  {journal} {\bibinfo  {journal} {Journal of the American Chemical Society}\ }\textbf {\bibinfo {volume} {146}},\ \bibinfo {pages} {8858--8864} (\bibinfo {year} {2024})}\BibitemShut {NoStop}%
\bibitem [{\citenamefont {Rai}\ \emph {et~al.}(2020)\citenamefont {Rai}, \citenamefont {Gerhard}, \citenamefont {Sun}, \citenamefont {Holzer}, \citenamefont {Rep{\"a}n}, \citenamefont {Krsti{\'c}}, \citenamefont {Yang}, \citenamefont {Wegener}, \citenamefont {Rockstuhl},\ and\ \citenamefont {Wulfhekel}}]{Rai2020-uv}%
  \BibitemOpen
  \bibfield  {author} {\bibinfo {author} {\bibfnamefont {Vibhuti}\ \bibnamefont {Rai}}, \bibinfo {author} {\bibfnamefont {Lukas}\ \bibnamefont {Gerhard}}, \bibinfo {author} {\bibfnamefont {Qing}\ \bibnamefont {Sun}}, \bibinfo {author} {\bibfnamefont {Christof}\ \bibnamefont {Holzer}}, \bibinfo {author} {\bibfnamefont {Taavi}\ \bibnamefont {Rep{\"a}n}}, \bibinfo {author} {\bibfnamefont {Marjan}\ \bibnamefont {Krsti{\'c}}}, \bibinfo {author} {\bibfnamefont {Liang}\ \bibnamefont {Yang}}, \bibinfo {author} {\bibfnamefont {Martin}\ \bibnamefont {Wegener}}, \bibinfo {author} {\bibfnamefont {Carsten}\ \bibnamefont {Rockstuhl}}, \ and\ \bibinfo {author} {\bibfnamefont {Wulf}\ \bibnamefont {Wulfhekel}},\ }\bibfield  {title} {\enquote {\bibinfo {title} {Boosting light emission from single hydrogen phthalocyanine molecules by charging},}\ }\href@noop {} {\bibfield  {journal} {\bibinfo  {journal} {Nano Lett.}\ }\textbf {\bibinfo {volume} {20}},\ \bibinfo {pages} {7600--7605} (\bibinfo {year} {2020})}\BibitemShut
  {NoStop}%
\bibitem [{\citenamefont {Kong}\ \emph {et~al.}(2022)\citenamefont {Kong}, \citenamefont {Tian}, \citenamefont {Zhang}, \citenamefont {Zhang}, \citenamefont {Chen}, \citenamefont {Yu}, \citenamefont {Jing}, \citenamefont {Gao}, \citenamefont {Luo}, \citenamefont {Yang}, \citenamefont {Dong},\ and\ \citenamefont {Hou}}]{Kong2022-tz}%
  \BibitemOpen
  \bibfield  {author} {\bibinfo {author} {\bibfnamefont {Fan-Fang}\ \bibnamefont {Kong}}, \bibinfo {author} {\bibfnamefont {Xiao-Jun}\ \bibnamefont {Tian}}, \bibinfo {author} {\bibfnamefont {Yang}\ \bibnamefont {Zhang}}, \bibinfo {author} {\bibfnamefont {Yao}\ \bibnamefont {Zhang}}, \bibinfo {author} {\bibfnamefont {Gong}\ \bibnamefont {Chen}}, \bibinfo {author} {\bibfnamefont {Yun-Jie}\ \bibnamefont {Yu}}, \bibinfo {author} {\bibfnamefont {Shi-Hao}\ \bibnamefont {Jing}}, \bibinfo {author} {\bibfnamefont {Hong-Ying}\ \bibnamefont {Gao}}, \bibinfo {author} {\bibfnamefont {Yi}~\bibnamefont {Luo}}, \bibinfo {author} {\bibfnamefont {Jin-Long}\ \bibnamefont {Yang}}, \bibinfo {author} {\bibfnamefont {Zhen-Chao}\ \bibnamefont {Dong}}, \ and\ \bibinfo {author} {\bibfnamefont {J~G}\ \bibnamefont {Hou}},\ }\bibfield  {title} {\enquote {\bibinfo {title} {Wavelike electronic energy transfer in donor--acceptor molecular systems through quantum coherence},}\ }\href@noop {} {\bibfield  {journal} {\bibinfo  {journal} {Nature
  Nanotechnology}\ }\textbf {\bibinfo {volume} {17}},\ \bibinfo {pages} {729--736} (\bibinfo {year} {2022})}\BibitemShut {NoStop}%
\bibitem [{\citenamefont {Imada}\ \emph {et~al.}(2017)\citenamefont {Imada}, \citenamefont {Miwa}, \citenamefont {Imai-Imada}, \citenamefont {Kawahara}, \citenamefont {Kimura},\ and\ \citenamefont {Kim}}]{Imada2017}%
  \BibitemOpen
  \bibfield  {author} {\bibinfo {author} {\bibfnamefont {Hiroshi}\ \bibnamefont {Imada}}, \bibinfo {author} {\bibfnamefont {Kuniyuki}\ \bibnamefont {Miwa}}, \bibinfo {author} {\bibfnamefont {Miyabi}\ \bibnamefont {Imai-Imada}}, \bibinfo {author} {\bibfnamefont {Shota}\ \bibnamefont {Kawahara}}, \bibinfo {author} {\bibfnamefont {Kensuke}\ \bibnamefont {Kimura}}, \ and\ \bibinfo {author} {\bibfnamefont {Yousoo}\ \bibnamefont {Kim}},\ }\bibfield  {title} {\enquote {\bibinfo {title} {Single-molecule investigation of energy dynamics in a coupled plasmon-exciton system},}\ }\href@noop {} {\bibfield  {journal} {\bibinfo  {journal} {Physical Review Letters}\ }\textbf {\bibinfo {volume} {119}} (\bibinfo {year} {2017})}\BibitemShut {NoStop}%
\bibitem [{\citenamefont {Zhang}\ \emph {et~al.}(2017{\natexlab{a}})\citenamefont {Zhang}, \citenamefont {Meng}, \citenamefont {Zhang}, \citenamefont {Luo}, \citenamefont {Yu}, \citenamefont {Yang}, \citenamefont {Zhang}, \citenamefont {Esteban}, \citenamefont {Aizpurua}, \citenamefont {Luo}, \citenamefont {Yang}, \citenamefont {Dong},\ and\ \citenamefont {Hou}}]{Zhang2017-yao}%
  \BibitemOpen
  \bibfield  {author} {\bibinfo {author} {\bibfnamefont {Yao}\ \bibnamefont {Zhang}}, \bibinfo {author} {\bibfnamefont {Qiu-Shi}\ \bibnamefont {Meng}}, \bibinfo {author} {\bibfnamefont {Li}~\bibnamefont {Zhang}}, \bibinfo {author} {\bibfnamefont {Yang}\ \bibnamefont {Luo}}, \bibinfo {author} {\bibfnamefont {Yun-Jie}\ \bibnamefont {Yu}}, \bibinfo {author} {\bibfnamefont {Ben}\ \bibnamefont {Yang}}, \bibinfo {author} {\bibfnamefont {Yang}\ \bibnamefont {Zhang}}, \bibinfo {author} {\bibfnamefont {Ruben}\ \bibnamefont {Esteban}}, \bibinfo {author} {\bibfnamefont {Javier}\ \bibnamefont {Aizpurua}}, \bibinfo {author} {\bibfnamefont {Yi}~\bibnamefont {Luo}}, \bibinfo {author} {\bibfnamefont {Jin-Long}\ \bibnamefont {Yang}}, \bibinfo {author} {\bibfnamefont {Zhen-Chao}\ \bibnamefont {Dong}}, \ and\ \bibinfo {author} {\bibfnamefont {J~G}\ \bibnamefont {Hou}},\ }\bibfield  {title} {\enquote {\bibinfo {title} {Sub-nanometre control of the coherent interaction between a single molecule and a plasmonic nanocavity},}\
  }\href@noop {} {\bibfield  {journal} {\bibinfo  {journal} {Nature Communications}\ }\textbf {\bibinfo {volume} {8}},\ \bibinfo {pages} {15225} (\bibinfo {year} {2017}{\natexlab{a}})}\BibitemShut {NoStop}%
\bibitem [{\citenamefont {Doppagne}\ \emph {et~al.}(2017)\citenamefont {Doppagne}, \citenamefont {Chong}, \citenamefont {Lorchat}, \citenamefont {Berciaud}, \citenamefont {Romeo}, \citenamefont {Bulou}, \citenamefont {Boeglin}, \citenamefont {Scheurer},\ and\ \citenamefont {Schull}}]{Doppagne2017-cq}%
  \BibitemOpen
  \bibfield  {author} {\bibinfo {author} {\bibfnamefont {Benjamin}\ \bibnamefont {Doppagne}}, \bibinfo {author} {\bibfnamefont {Michael~C.}\ \bibnamefont {Chong}}, \bibinfo {author} {\bibfnamefont {Etienne}\ \bibnamefont {Lorchat}}, \bibinfo {author} {\bibfnamefont {St\'ephane}\ \bibnamefont {Berciaud}}, \bibinfo {author} {\bibfnamefont {Michelangelo}\ \bibnamefont {Romeo}}, \bibinfo {author} {\bibfnamefont {Herv\'e}\ \bibnamefont {Bulou}}, \bibinfo {author} {\bibfnamefont {Alex}\ \bibnamefont {Boeglin}}, \bibinfo {author} {\bibfnamefont {Fabrice}\ \bibnamefont {Scheurer}}, \ and\ \bibinfo {author} {\bibfnamefont {Guillaume}\ \bibnamefont {Schull}},\ }\bibfield  {title} {\enquote {\bibinfo {title} {Vibronic spectroscopy with submolecular resolution from stm-induced electroluminescence},}\ }\href@noop {} {\bibfield  {journal} {\bibinfo  {journal} {Phys. Rev. Lett.}\ }\textbf {\bibinfo {volume} {118}},\ \bibinfo {pages} {127401} (\bibinfo {year} {2017})}\BibitemShut {NoStop}%
\bibitem [{\citenamefont {Chen}\ \emph {et~al.}(2019)\citenamefont {Chen}, \citenamefont {Luo}, \citenamefont {Gao}, \citenamefont {Jiang}, \citenamefont {Yu}, \citenamefont {Zhang}, \citenamefont {Zhang}, \citenamefont {Li}, \citenamefont {Zhang},\ and\ \citenamefont {Dong}}]{chen2019spin}%
  \BibitemOpen
  \bibfield  {author} {\bibinfo {author} {\bibfnamefont {Gong}\ \bibnamefont {Chen}}, \bibinfo {author} {\bibfnamefont {Yang}\ \bibnamefont {Luo}}, \bibinfo {author} {\bibfnamefont {Hongying}\ \bibnamefont {Gao}}, \bibinfo {author} {\bibfnamefont {Jun}\ \bibnamefont {Jiang}}, \bibinfo {author} {\bibfnamefont {Yunjie}\ \bibnamefont {Yu}}, \bibinfo {author} {\bibfnamefont {Li}~\bibnamefont {Zhang}}, \bibinfo {author} {\bibfnamefont {Yang}\ \bibnamefont {Zhang}}, \bibinfo {author} {\bibfnamefont {Xiaoguang}\ \bibnamefont {Li}}, \bibinfo {author} {\bibfnamefont {Zhenyu}\ \bibnamefont {Zhang}}, \ and\ \bibinfo {author} {\bibfnamefont {Zhenchao}\ \bibnamefont {Dong}},\ }\bibfield  {title} {\enquote {\bibinfo {title} {Spin-triplet-mediated up-conversion and crossover behavior in single-molecule electroluminescence},}\ }\href@noop {} {\bibfield  {journal} {\bibinfo  {journal} {Physical Review Letters}\ }\textbf {\bibinfo {volume} {122}},\ \bibinfo {pages} {177401} (\bibinfo {year} {2019})}\BibitemShut {NoStop}%
\bibitem [{\citenamefont {Luo}\ \emph {et~al.}(2024)\citenamefont {Luo}, \citenamefont {Kong}, \citenamefont {Tian}, \citenamefont {Yu}, \citenamefont {Jing}, \citenamefont {Zhang}, \citenamefont {Chen}, \citenamefont {Zhang}, \citenamefont {Zhang}, \citenamefont {Li} \emph {et~al.}}]{luo2024anomalously}%
  \BibitemOpen
  \bibfield  {author} {\bibinfo {author} {\bibfnamefont {Yang}\ \bibnamefont {Luo}}, \bibinfo {author} {\bibfnamefont {Fan-Fang}\ \bibnamefont {Kong}}, \bibinfo {author} {\bibfnamefont {Xiao-Jun}\ \bibnamefont {Tian}}, \bibinfo {author} {\bibfnamefont {Yun-Jie}\ \bibnamefont {Yu}}, \bibinfo {author} {\bibfnamefont {Shi-Hao}\ \bibnamefont {Jing}}, \bibinfo {author} {\bibfnamefont {Chao}\ \bibnamefont {Zhang}}, \bibinfo {author} {\bibfnamefont {Gong}\ \bibnamefont {Chen}}, \bibinfo {author} {\bibfnamefont {Yang}\ \bibnamefont {Zhang}}, \bibinfo {author} {\bibfnamefont {Yao}\ \bibnamefont {Zhang}}, \bibinfo {author} {\bibfnamefont {Xiao-Guang}\ \bibnamefont {Li}},  \emph {et~al.},\ }\bibfield  {title} {\enquote {\bibinfo {title} {Anomalously bright single-molecule upconversion electroluminescence},}\ }\href@noop {} {\bibfield  {journal} {\bibinfo  {journal} {Nature Communications}\ }\textbf {\bibinfo {volume} {15}},\ \bibinfo {pages} {1677} (\bibinfo {year} {2024})}\BibitemShut {NoStop}%
\bibitem [{\citenamefont {Zhang}\ \emph {et~al.}(2017{\natexlab{b}})\citenamefont {Zhang}, \citenamefont {Yu}, \citenamefont {Chen}, \citenamefont {Luo}, \citenamefont {Yang}, \citenamefont {Kong}, \citenamefont {Chen}, \citenamefont {Zhang}, \citenamefont {Zhang}, \citenamefont {Luo} \emph {et~al.}}]{zhang2017electrically}%
  \BibitemOpen
  \bibfield  {author} {\bibinfo {author} {\bibfnamefont {Li}~\bibnamefont {Zhang}}, \bibinfo {author} {\bibfnamefont {Yun-Jie}\ \bibnamefont {Yu}}, \bibinfo {author} {\bibfnamefont {Liu-Guo}\ \bibnamefont {Chen}}, \bibinfo {author} {\bibfnamefont {Yang}\ \bibnamefont {Luo}}, \bibinfo {author} {\bibfnamefont {Ben}\ \bibnamefont {Yang}}, \bibinfo {author} {\bibfnamefont {Fan-Fang}\ \bibnamefont {Kong}}, \bibinfo {author} {\bibfnamefont {Gong}\ \bibnamefont {Chen}}, \bibinfo {author} {\bibfnamefont {Yang}\ \bibnamefont {Zhang}}, \bibinfo {author} {\bibfnamefont {Qiang}\ \bibnamefont {Zhang}}, \bibinfo {author} {\bibfnamefont {Yi}~\bibnamefont {Luo}},  \emph {et~al.},\ }\bibfield  {title} {\enquote {\bibinfo {title} {Electrically driven single-photon emission from an isolated single molecule},}\ }\href@noop {} {\bibfield  {journal} {\bibinfo  {journal} {Nature communications}\ }\textbf {\bibinfo {volume} {8}},\ \bibinfo {pages} {580} (\bibinfo {year} {2017}{\natexlab{b}})}\BibitemShut {NoStop}%
\bibitem [{\citenamefont {Luo}\ \emph {et~al.}(2019)\citenamefont {Luo}, \citenamefont {Chen}, \citenamefont {Zhang}, \citenamefont {Zhang}, \citenamefont {Yu}, \citenamefont {Kong}, \citenamefont {Tian}, \citenamefont {Zhang}, \citenamefont {Shan}, \citenamefont {Luo} \emph {et~al.}}]{luo2019electrically}%
  \BibitemOpen
  \bibfield  {author} {\bibinfo {author} {\bibfnamefont {Yang}\ \bibnamefont {Luo}}, \bibinfo {author} {\bibfnamefont {Gong}\ \bibnamefont {Chen}}, \bibinfo {author} {\bibfnamefont {Yang}\ \bibnamefont {Zhang}}, \bibinfo {author} {\bibfnamefont {Li}~\bibnamefont {Zhang}}, \bibinfo {author} {\bibfnamefont {Yunjie}\ \bibnamefont {Yu}}, \bibinfo {author} {\bibfnamefont {Fanfang}\ \bibnamefont {Kong}}, \bibinfo {author} {\bibfnamefont {Xiaojun}\ \bibnamefont {Tian}}, \bibinfo {author} {\bibfnamefont {Yao}\ \bibnamefont {Zhang}}, \bibinfo {author} {\bibfnamefont {Chongxin}\ \bibnamefont {Shan}}, \bibinfo {author} {\bibfnamefont {Yi}~\bibnamefont {Luo}},  \emph {et~al.},\ }\bibfield  {title} {\enquote {\bibinfo {title} {Electrically driven single-photon superradiance from molecular chains in a plasmonic nanocavity},}\ }\href@noop {} {\bibfield  {journal} {\bibinfo  {journal} {Physical Review Letters}\ }\textbf {\bibinfo {volume} {122}},\ \bibinfo {pages} {233901} (\bibinfo {year} {2019})}\BibitemShut {NoStop}%
\bibitem [{\citenamefont {Ploigt}\ \emph {et~al.}(2007)\citenamefont {Ploigt}, \citenamefont {Brun}, \citenamefont {Pivetta}, \citenamefont {Patthey},\ and\ \citenamefont {Schneider}}]{ploigt2007local}%
  \BibitemOpen
  \bibfield  {author} {\bibinfo {author} {\bibfnamefont {Hans-Christoph}\ \bibnamefont {Ploigt}}, \bibinfo {author} {\bibfnamefont {Christophe}\ \bibnamefont {Brun}}, \bibinfo {author} {\bibfnamefont {Marina}\ \bibnamefont {Pivetta}}, \bibinfo {author} {\bibfnamefont {Fran{\c{c}}ois}\ \bibnamefont {Patthey}}, \ and\ \bibinfo {author} {\bibfnamefont {Wolf-Dieter}\ \bibnamefont {Schneider}},\ }\bibfield  {title} {\enquote {\bibinfo {title} {Local work function changes determined by field emission resonances:{{NaCl}/{Ag}(100)}},}\ }\href@noop {} {\bibfield  {journal} {\bibinfo  {journal} {Physical Review B—Condensed Matter and Materials Physics}\ }\textbf {\bibinfo {volume} {76}},\ \bibinfo {pages} {195404} (\bibinfo {year} {2007})}\BibitemShut {NoStop}%
\bibitem [{\citenamefont {Banerjee}\ \emph {et~al.}(2025)\citenamefont {Banerjee}, \citenamefont {Ide}, \citenamefont {Lu}, \citenamefont {Berndt},\ and\ \citenamefont {Weismann}}]{Banerjee2025-ze}%
  \BibitemOpen
  \bibfield  {author} {\bibinfo {author} {\bibfnamefont {Arnab}\ \bibnamefont {Banerjee}}, \bibinfo {author} {\bibfnamefont {Niklas}\ \bibnamefont {Ide}}, \bibinfo {author} {\bibfnamefont {Yan}\ \bibnamefont {Lu}}, \bibinfo {author} {\bibfnamefont {Richard}\ \bibnamefont {Berndt}}, \ and\ \bibinfo {author} {\bibfnamefont {Alexander}\ \bibnamefont {Weismann}},\ }\bibfield  {title} {\enquote {\bibinfo {title} {{Adsorption-Site-} and {Orientation-Dependent} magnetism of a molecular switch on pb(100)},}\ }\href@noop {} {\bibfield  {journal} {\bibinfo  {journal} {ACS Nano}\ }\textbf {\bibinfo {volume} {19}},\ \bibinfo {pages} {7231--7238} (\bibinfo {year} {2025})}\BibitemShut {NoStop}%
\bibitem [{\citenamefont {Wang}\ \emph {et~al.}(2009)\citenamefont {Wang}, \citenamefont {Kr{\"o}ger}, \citenamefont {Berndt},\ and\ \citenamefont {Hofer}}]{Wang2009-bo}%
  \BibitemOpen
  \bibfield  {author} {\bibinfo {author} {\bibfnamefont {Yongfeng}\ \bibnamefont {Wang}}, \bibinfo {author} {\bibfnamefont {J{\"o}rg}\ \bibnamefont {Kr{\"o}ger}}, \bibinfo {author} {\bibfnamefont {Richard}\ \bibnamefont {Berndt}}, \ and\ \bibinfo {author} {\bibfnamefont {Werner~A}\ \bibnamefont {Hofer}},\ }\bibfield  {title} {\enquote {\bibinfo {title} {Pushing and pulling a sn ion through an adsorbed phthalocyanine molecule},}\ }\href@noop {} {\bibfield  {journal} {\bibinfo  {journal} {J. Am. Chem. Soc.}\ }\textbf {\bibinfo {volume} {131}},\ \bibinfo {pages} {3639--3643} (\bibinfo {year} {2009})}\BibitemShut {NoStop}%
\bibitem [{\citenamefont {Reecht}\ \emph {et~al.}(2019)\citenamefont {Reecht}, \citenamefont {Krane}, \citenamefont {Lotze},\ and\ \citenamefont {Franke}}]{Reecht2019-qn}%
  \BibitemOpen
  \bibfield  {author} {\bibinfo {author} {\bibfnamefont {Ga{\"e}l}\ \bibnamefont {Reecht}}, \bibinfo {author} {\bibfnamefont {Nils}\ \bibnamefont {Krane}}, \bibinfo {author} {\bibfnamefont {Christian}\ \bibnamefont {Lotze}}, \ and\ \bibinfo {author} {\bibfnamefont {Katharina~J}\ \bibnamefont {Franke}},\ }\bibfield  {title} {\enquote {\bibinfo {title} {{$\pi$-Radical} formation by pyrrolic {H} abstraction of phthalocyanine molecules on molybdenum disulfide},}\ }\href@noop {} {\bibfield  {journal} {\bibinfo  {journal} {ACS Nano}\ }\textbf {\bibinfo {volume} {13}},\ \bibinfo {pages} {7031--7035} (\bibinfo {year} {2019})}\BibitemShut {NoStop}%
\bibitem [{\citenamefont {N{\'e}el}\ \emph {et~al.}(2016)\citenamefont {N{\'e}el}, \citenamefont {Lattelais}, \citenamefont {Bocquet},\ and\ \citenamefont {Kr{\"o}ger}}]{Neel2016-gm}%
  \BibitemOpen
  \bibfield  {author} {\bibinfo {author} {\bibfnamefont {Nicolas}\ \bibnamefont {N{\'e}el}}, \bibinfo {author} {\bibfnamefont {Marie}\ \bibnamefont {Lattelais}}, \bibinfo {author} {\bibfnamefont {Marie-Laure}\ \bibnamefont {Bocquet}}, \ and\ \bibinfo {author} {\bibfnamefont {J{\"o}rg}\ \bibnamefont {Kr{\"o}ger}},\ }\bibfield  {title} {\enquote {\bibinfo {title} {Depopulation of {Single-Phthalocyanine} molecular orbitals upon {Pyrrolic-Hydrogen} abstraction on graphene},}\ }\href@noop {} {\bibfield  {journal} {\bibinfo  {journal} {ACS Nano}\ }\textbf {\bibinfo {volume} {10}},\ \bibinfo {pages} {2010--2016} (\bibinfo {year} {2016})}\BibitemShut {NoStop}%
\bibitem [{\citenamefont {Miwa}\ \emph {et~al.}(2013)\citenamefont {Miwa}, \citenamefont {Sakaue},\ and\ \citenamefont {Kasai}}]{Miwa2013}%
  \BibitemOpen
  \bibfield  {author} {\bibinfo {author} {\bibfnamefont {Kuniyuki}\ \bibnamefont {Miwa}}, \bibinfo {author} {\bibfnamefont {Mamoru}\ \bibnamefont {Sakaue}}, \ and\ \bibinfo {author} {\bibfnamefont {Hideaki}\ \bibnamefont {Kasai}},\ }\bibfield  {title} {\enquote {\bibinfo {title} {Effects of interference between energy absorption processes of molecule and surface plasmons on light emission induced by scanning tunneling microscopy},}\ }\href@noop {} {\bibfield  {journal} {\bibinfo  {journal} {Journal of the Physical Society of Japan}\ }\textbf {\bibinfo {volume} {82}},\ \bibinfo {pages} {124707} (\bibinfo {year} {2013})}\BibitemShut {NoStop}%
\end{thebibliography}%

\section*{Methods}
\textbf{Sample preparation.} 
The Ag(111) surface was prepared through repeated cycles of Ar$^+$ ion sputtering followed by annealing. NaCl was deposited onto Ag(111) at room temperature using a Knudsen cell evaporator heated to 813 K. ZnPc and SnPc molecules were subsequently deposited onto the NaCl-covered Ag(111) at 7 K, with the evaporator maintained at 690 K. STM tips were fabricated by electrochemical etching of Ag wires and further conditioned by controlled indentation and voltage pulsing on the Ag(111) surface to get desired plasmonic active status.

\textbf{STM measurements.} All experiments were carried out using a low-temperature STM (Unisoku-1400) operating at 7 K under ultrahigh vacuum conditions. STM topographs were obtained in constant-current mode, where the tip height was regulated by a feedback loop to maintain a constant tunneling current. Differential conductance (d$I$/d$V$) spectra were recorded with the feedback loop open, employing a standard lock-in technique with a bias modulation of 30 mV at 403 Hz.

\textbf{STML measurements.}  STML Spectra were recorded with a  spectrometer (with CCD) equipped with a 150 grooves / $mm$ grating and 150 $\mu m$ slit. All spectra shown in this paper are presented without correction for the wavelength-dependent sensitivity of the photon detection system.

\end{document}